# Asymmetric phenomenon of flow and heat transfer in charging process of thermal energy storage based on an entire domain model

Xuchen Ying[a], Weijia Huang[a, *], Wenhua Liu[a,b], Guiliang Liu[a], Jun Li[a], Mo Yang[c,a,*]

[a] *School of Energy and Power Engineering, University of Shanghai for Science and Technology, Shanghai 200093, China*

[b] *Department of Mechanical and Aerospace Engineering, University of Missouri, Columbia, MO 65211, USA*

[c] *Shanghai Jian Qiao University, Shanghai 201306, China*

**Abstract**

Phase change energy storage is getting increasing attention as a representative technology to achieve carbon neutrality. The phase change process exists typical phenomenon of asymmetry that affects the energy storage performance. However, the mechanism of asymmetry is currently lack of elaboration because the half domain model is always used to simplify the numerical simulation and avoid the appearance of asymmetry. In this study, the entire domain model and boundary conditions were adopted and numerically simulated for the melting process of paraffin wax, i.e., the charging process of energy storage. The nonlinear dynamics method was applied to explain the asymmetric flow and heat transfer phenomenon. The charging model was verified at first, and the pear-shaped contour maps of temperature distribution, flow pattern, and liquid fraction were obtained. Then, three important Rayleigh numbers were found according to the stability of flow and heat transfer. The three-stage characteristic based on charging speed was proposed for charging process and was explained by thermal conduction or natural convection. The asymmetric phenomenon was elaborated on the cause of formation, change mechanism, and effect evaluation. Results show that natural convection accounts for the multiple solutions (including the asymmetric ones) of charging process. It is also pointed out that asymmetric solutions can exist under symmetric geometry structure. To accurately solve the charging process, it is necessary to use the entire domain model. Both the charging speed and energy storage capacity have a positive correlation with asymmetry. Thus, a potential idea for energy storage application was proposed to increase the charging speed; that is, adding thermal disturbance during the melting process to destroy flow stability and to enhance convective heat transfer. The research methods and findings will support the development of energy storage and numerical investigation in other related areas.

\* Corresponding author at: School of Energy and Power Engineering, University of Shanghai for Science and Technology, Shanghai 200093, China.

*E-mail address:* huang.weijia@foxmail.com (Weijia Huang).

\* Corresponding author at: Shanghai Jian Qiao University, Shanghai 201306, China; School of Energy and Power Engineering, University of Shanghai for Science and Technology, Shanghai 200093, China

*E-mail address:* yangm@usst.edu.cn (Mo Yang).

# 1. Introduction

Along with the growth of global economy, energy consumption in the world has increased exponentially in recent years, although there has been a negative impact from the terrible pandemic COVID-19. The growing energy demand suggests that the world is currently facing a severe challenge concerning energy shortage. The coordination of energy in time and space is critical to addressing the energy problem. As is generally understood, "time" means that energy usage varies with time, especially solar energy during the day and night [1]. Similarly, "space" means that energy usage and production mismatches in geographical imbalance. These problems suggest that it is necessary to use energy storage system to keep high efficiency running for renewable energy applications and ultimate achievement of sustainable development with carbon neutrality. Hence, increasing attention has been focused on the latent heat thermal energy storage (LHTES) [2], i.e., phase change energy storage, which is found in many applications, such as solar still [3], power generation [2,4], and ventilation systems in buildings [5,6].

Efforts in LHTES study can be divided into three parts [7]: property of phase change material (PCM), heat transfer enhancement structure of thermal energy storage unit, and theoretical mechanism of charging and discharging processes. Experimental studies aim to discover PCM with better thermal conductivity, stability, and thermal energy storage density. Shao et al. [8,9] tested the melting and solidification behaviors of various sugar alcohols as PCM. Paraffin wax is the most typically commercial organic PCM, it is non-corrosive to most equipment and pipeline, and it does not suffer from sub-cooling with over 1500 stable cycles [10]. In terms of heat transfer enhancement structure. Pizzolato et al. [11] presented a unique heat transfer intensification design for shell-and-tube LHTES units. Abbassi et al. [12] proposed a mathematical model for the transient simulation of the large-size thermal energy storage tank of ice. As for the model geometry, various sizes of the thermal energy storage tank and the storage unit in applications can be found [13]. Choi et al. [14] experimentally explored the method of bubble-driven flow by two heat pipes with the shape of 40×60 mm rectangular. Zheng [15] et al. used a 30 mm two-dimensional circular model to study the application in the concrete structure design of LHTES unit. Shahsavar et al. [14] numerically investigated the melting and solidification processes of the double-pipe PCM heat exchanger, and results show this type of structure significantly impacts on the charging speed. These studies about PCM and heat transfer structure indicate that paraffin wax is the most widely applicable and commercial material in the field of energy storage, and the study of energy storage units is widely accepted. In the theoretical studies of physical mechanism, papers focus on the correlation between dimensionless parameters and liquid fraction. Gao et al. [16] proposed that the Rayleigh number (Ra) significantly impacts the transition point between heat conduction and convection. The follow-up work [17] focused on the effective thermal conductivity of the thermal energy storage unit and found out it correlates stronger with Ra than Stefan number (Ste). Farsani et al. [18] found that melting rate and heat transfer increased by 8% in the case of rotation with the control of Taylor number and *Ra*. These theoretical articles about phase changing behavior focus on the study of Ra and Ste, which shows that these two dimensionless numbers affect the state of the charging process.

Although many numerical simulations have been conducted, the half domain model is currently used in most literatures to predict the charging process due to the consideration on symmetric physical structure of model and boundary conditions. Archibold et al. [19] investigated PCM in filled spherical shell and generalized the correlation for liquid mass fraction and Nusselt number. Soni et al. [20] analyzed the constrained melting of PCM based on an entire domain model and observed thermal plumes and multiple counter-rotating vortices in the capsule, but they did not consider the nonlinearity; thus, only half of the entire domain model worked. When conducting the experiments for charging process, Tan [21] noticed the asymmetric phenomenon appeared but attributed it to some external disturbance. In numerical calculations the model and result are often consider as strictly symmetric. Karthikeyan et al. [22] obtained the noticeable asymmetric numerical results but did not give the explanation, though they are different with the symmetric results in other literatures [15,16]. Multiple solutions, including the asymmetric and symmetric solutions, are one of the characteristics in nonlinear phenomenon, but the inertial thinking that one symmetric solution for the flow in a symmetric structure may make researchers ignore the asymmetric results.

As for the problem of the solid-liquid phase transition during the charging process, few people paid attention to the nonlinear phenomenon. Indeed, nonlinearity and nonlinear phenomenon is ubiquitous in nature [23]. The nonlinear phenomenon and bifurcation occur when Navier-Stokes equation exists multiple stable solutions, which has been proved in mathematic [24]. More and more researchers have paid attention to the nonlinear characteristics of the physical system involving convection and have noticed the phenomenon of multiple solutions. Inoue et al. [25] studied the chaotic itinerancy, a typically nonlinear phenomenon found in optical turbulence, to improve the design of the neural network. Gavara [26] reported the asymmetric flow and heat transfer of numerical results in the symmetric geometry and heating channel. Few studies have linked nonlinearity with the asymmetric phenomenon during the charging process of LHTES.

The understanding of the nonlinearity, e.g., asymmetric phenomenon, in system is critical to both preventing nonlinear influence (when system stability is needed) and elevating engineering nonlinearity (when system performance is desired) [27–29]. One typical application of nonlinearity is the "Sequential start" method to reduce the thermal deviation resulting from the nonlinearity in boiler control. Ma et al. [30] focused on the asymmetric phenomenon of burning process with its effect on energy efficiency and set up that method. Some scholars have also explored the nonlinear impact of flow and heat transfer in different scenes. Li et al. [31] discussed the flow and mass transfer bifurcation in the cylindrical and semi-cylindrical models. Sung and Kim [32] also provided evidence of bifurcation for the phase change process. These explorations are extraordinary and have inspired people's understanding of the complex physical system. However, the understanding of nonlinearity in engineering of energy storage is far from enough. Mahdi et al. [33] investigated the charging process with various PCM arrangements. Few of these widely cited experimental and numerical studies have linked nonlinearity with asymmetric phenomenon but have used the half domain to avoid the discussion of the asymmetric phenomenon. Therefore, studying nonlinearity will help explore the theoretical mechanism of LHTES.

To elaborate the mechanism of asymmetric phenomenon for flow and heat transfer during the charging process of phase change energy storage, the present study numerically investigated paraffin wax's melting process around an isothermal and horizontal heater. The entire domain model with symmetric boundary conditions were adopted, and the nonlinear dynamics method was applied to explain the asymmetric phenomenon. The commercial software FLUENT was used as the solver to calculate the governing equations. A homemade Python code was developed to analyze the nonlinear characteristic of the data. The mechanism of nonlinear charging process was explored, the stability of flow was investigated, and the effects of asymmetric phenomenon on energy storage application were discussed based on the findings.

## 2. Physical model and basic equation

### 2.1 Mathematical model

Considering the influence of the left and right vortexes squeezing the spherical capsule [20], this study used the square cavity model as the computational domain The physical model is a 100 mm square cavity with a horizontal cylinder heater, as shown in **Fig. 1**, to represent the PCM unit in the LHTES. The energy storage unit has been widely used for exploring and improving the methods to enhance the performance of LHTES application [17]. Heat transfer fluid is inside the heater; the other space around the cylinder is filled with PCM, i.e., paraffin wax in this study.

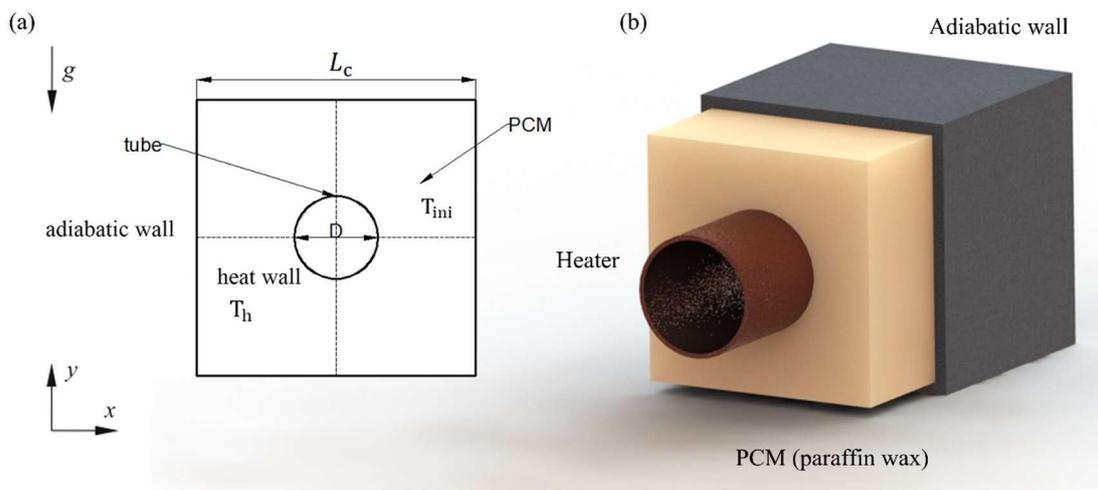

**Fig. 1** The square enclosure with horizontal cylinder: (a) computational domain; (b) physical model

The charging process can be governed by the fundamental equations of continuity, momentum, and energy. Besides, the following assumptions are used to solve these fundamental equations [34].

1. The Boussinesq approximation is valid.
2. Viscous dissipation of PCM is neglected.
3. The outer wall of the enclosure is isothermal.
4. The properties of PCM are homogeneous and isotropic.
5. The flow of the melting process is laminar and two-dimensional.
6. The temperature of the heater wall is constant for each case.

The enthalpy-porosity method is used to avoid considering the solid-liquid interface during the melting process. Instead, a variable called the liquid fraction ($\alpha$) is used to get the temperature solution. The primary governing equations of the charging process model using the enthalpy-porosity method are given by **Eqs. (1)-(3)**.

Continuity Equation,

$$\nabla \cdot (\rho \boldsymbol{v}) = 0. \tag{1}$$

Momentum Equation,

$$\frac{\partial(\rho \boldsymbol{v})}{\partial t} + \nabla \cdot (\rho \boldsymbol{v}\boldsymbol{v}) = \nabla \cdot (\nu \rho \cdot \nabla \boldsymbol{v}) - \nabla p + \boldsymbol{S}_\mathrm{m} + \boldsymbol{S}_\mathrm{b}. \tag{2}$$

Energy Equation,

$$\frac{\partial(\rho h)}{\partial t} + \nabla \cdot (\rho \boldsymbol{v} h) = \nabla \cdot (\lambda \cdot \nabla T), \tag{3}$$

where $\rho$ is the density, $\boldsymbol{v}$ is the velocity vector, $p$ is the pressure, $h$ is the specific enthalpy, and $\lambda$ is the thermal conductivity of PCM.

The specific enthalpy of PCM in **Eq. (3)** is computed as the sum of sensible heat ($h_\mathrm{sen}$) and latent heat content ($\Delta h$). The sensible heat can be achieved by **Eq. (4)**,

$$h_\mathrm{sen} = h_\mathrm{ref} + \int_{T_\mathrm{ref}}^{T} c_\mathrm{l} dT. \tag{4}$$

Further, the source term $\boldsymbol{S}_m$ in the momentum equation is called Darcy type source term and the rest source term $\boldsymbol{S}_b$ is called buoyancy term, which only exists in the *Y*-axis momentum equation [35]. Darcy type source $\boldsymbol{S}_m$ is defined by **Eq. (5)**,

$$S_{\mathrm{m}} = \frac{(1-\alpha)^2}{(\alpha^3 + \varepsilon)} A_{\mathrm{mush}} \boldsymbol{v}, \tag{5}$$

where $A_{\mathrm{mush}}$ represents the mush zone in the calculation region, a phase transition state between solid and liquid, and is generally regarded as a significant constant value. $\varepsilon$ is a small number used to prevent this equation from dividing to zero when the region of PCM is solid and the liquid fraction $\alpha$ is equal to zero [36]. In this study, $A_{\mathrm{mush}}$ is $1 \times 10^5$ kg/(m$^3 \cdot$ s) and $\varepsilon$ is set as 0.001.

Liquid fraction $\alpha$ is the most significant value to locate the interface in the mush zone, wherein $\alpha$ lies between 1 and 0. When the PCM is in the solid state, the liquid fraction is 0; whereas the PCM is completely melting to the liquid state, the liquid fraction turns out to 1.

With the concept of liquid fraction, the relationship between the temperature and the enthalpy can be described as **Eq. (6)**,

$$T = T_{\mathrm{s}} + \frac{\Delta h (T_{\mathrm{l}} - T_{\mathrm{s}})}{h_{\mathrm{f}}}. \tag{6}$$

Subsequently, $\alpha$ can be defined by **Eq. (7)**,

$$\alpha = \frac{\Delta h}{h_{\mathrm{f}}} = \begin{cases} 0, & T < T_{\mathrm{s}} \\ \frac{T - T_{\mathrm{s}}}{T_{\mathrm{l}} - T_{\mathrm{s}}}, & T_{\mathrm{s}} \leq T \leq T_{\mathrm{l}} \\ 1, & T > T_{\mathrm{l}} \end{cases}. \tag{7}$$

According to Boussinesq assumption, the buoyancy source term $\boldsymbol{S}_{\mathrm{b}}$ can be described by

$$\boldsymbol{S}_{\mathrm{b}} = \rho g \beta (T - T_{\mathrm{ref}}), \tag{8}$$

where $\beta$ is the volumetric thermal expansion coefficient of liquid, and $T_{\mathrm{ref}}$ is the reference temperature.

**2.2 Normalization**

The dimensionless procedure is customarily used to improve analysis accuracy during phase change process [37]. The dimensionless forms of the governing equations are written as follows.

Continuity Equation,

$$\nabla \cdot (\boldsymbol{V}) = 0. \tag{9}$$

Momentum Equation,

$$\frac{\partial (\boldsymbol{V})}{\partial Fo} + \nabla \cdot (\boldsymbol{V}\boldsymbol{V}) = Pr \cdot \nabla \cdot (\nabla \boldsymbol{V}) - \nabla P + \boldsymbol{S}_{\mathrm{M}} + \boldsymbol{S}_{\mathrm{B}}. \tag{10}$$

Energy Equation,

$$\frac{\partial (H)}{\partial Fo} + \nabla \cdot (\boldsymbol{V} H) = Ste \cdot \nabla \cdot \nabla \theta, \tag{11}$$

where $\theta = \frac{T-T_\mathrm{m}}{T_\mathrm{h}-T_\mathrm{m}}$, $H = \frac{h}{h_\mathrm{f}}$, $\boldsymbol{V} = \frac{\boldsymbol{v}\rho c_1 L_\mathrm{c}}{k_1}$, and $P = \frac{\rho L_\mathrm{c}^2 c_\mathrm{p}^2 p}{k^2}$.

The liquid fraction $\alpha$ is rewritten as

$$\alpha = \begin{cases} 0, & \theta < \theta_\mathrm{s} \\ H, & \theta_\mathrm{s} \leq \theta \leq \theta_\mathrm{l} \\ 1, & \theta > \theta_\mathrm{l} \end{cases}. \tag{12}$$

The Darcy type source term can be written by

$$S_\mathrm{M} = \frac{(1-\alpha)^2}{(\alpha^3+\varepsilon)} A_\mathrm{MUSH} \boldsymbol{V}, \tag{13}$$

where $A_\mathrm{MUSH} = \frac{c_1 A_\mathrm{mush} L_\mathrm{c}^2}{\lambda_1}$.

The Buoyancy source term can be written by

$$S_\mathrm{B} = Pr \cdot Ra. \tag{14}$$

The Fourier number ($Fo$) in **Eq. (11)** is defined as

$$Fo = \frac{\lambda_1 t}{\rho c_1 L_\mathrm{c}^2}. \tag{15}$$

The Prandtl number ($Pr$) in **Eq. (14)** is defined as

$$Pr = \frac{c_1 \nu \rho}{\lambda_1}. \tag{16}$$

In addition, $Ra$ and $Ste$ are used to characterize the melting process in this type of study. $Ra$ is the dimensionless parameter associated with the buoyancy-driven flow and is defined as

$$Ra = \frac{g\rho c_1 \Delta T \beta L_\mathrm{c}^3}{\nu \lambda_1}, \tag{17}$$

where $\nu$ is the kinematic viscosity of the liquid state of PCM, $\Delta T$ represents the temperature difference between the heater and melting PCM, and $L_c$ is the characteristic length and is the height of the central melting region in this study.

$Ste$ is a dimensionless ratio of sensible heat to latent heat, which is given by

$$Ste = \frac{c_1 \Delta T}{h_\mathrm{f}}. \tag{18}$$

This dimensionless parameter helps analyze the Stefan problem that covers the presented cases of this study. The combination of dimensionless number $Ste \cdot Fo$ is generally used for a characteristic time in analyzing melting process.

**2.3 Boundary and initial conditions**

Initially, when $t = 0$, the PCM is in the solid state, the cavity wall and all zone of PCM are at the same temperature that is lower than the melting temperature of PCM. The initial temperature ($T_\mathrm{ini}$) is maintained at 330 K and 320 K for temperature differences ($\Delta T$) of 10 K and 40 K, respectively. Besides, the temperature of the round wall of the horizontal cylinder where the heat transfer fluid flow inside keeps a constant value, i.e., negligible heat loss of the charging model is assumed. The no-slip boundary condition and adiabatic condition are also applied at the cavity wall. The temperature difference ($\Delta T$) for each case is controlled by the heating wall; according to **Eqs. (17) and (18)**, $Ra$ and $Ste$ are consequently controlled. Finally, the temperature of the heater ($T_h$) is maintained at 351 K and 381 K for temperature difference ($\Delta T$) of 10 K and 40 K, respectively.

## 3. Model validation and numerical simulation

To obtain a general conclusion, the present study selects a classical physical properties of paraffin wax, which can be found in **Table 1**. This kind of paraffin wax with low melting point is widely used in numerical simulations [22,38]. The specific heat storage capacity of different paraffin is similar. The phase change energy storage mainly depends on the latent heat of PCM, of which the value adopted in this work is 200 kJ/kg, which is the median level in low-temperature materials [39], such as paraffin, non-paraffin organics, inorganic compounds, eutectic compounds. Except for melting point and heat capacity, the low thermal conductivity is a disadvantage for PCM; thus, the paraffin wax with a relatively high thermal conductivity of 0.2 W/(m·K) was selected in this work [40]. For the given geometry model, the values of $Ra$ and $Ste$ correspond to the temperature difference $\Delta T$ and the dimensional length $L_c$. Hence, different cases of charging processes with different $Ra$ and $Ste$ values were designed.

Table 1 Physical properties of paraffin wax [41]

| Properties | Values | Units |
|---|---|---|
| Density, $\rho$ | 920 | kg/m$^3$ |
| Reference temperature, $T_{ref}$ | 341 | K |
| The volumetric thermal expansion coefficient of liquid, $\beta$ | 8.5×10$^{-4}$ | 1/K |
| Thermal conductivity (liquid), $\lambda_l$ | 0.20 | W/(m·K) |
| Thermal conductivity (solid), $\lambda_s$ | 0.45 | W/(m·K) |
| Latent heat of fusion, $h_f$ | 203.9 | kJ/kg |
| Specific heat (liquid), $c_l$ | 2.1 | kJ/(kg·K) |
| Specific heat (solid), $c_s$ | 2.6 | kJ/(kg·K) |
| Kinematic viscosity (liquid), $\nu$ | 0.01 | m$^2$/s |
| Melting temperature, $T_m$ | 343 | K |

The experimental data measured by Gau and Viskanta [42] was widely used to verify the numerical method. The experiment focuses on the melting process of natural convection in a rectangular enclosure. **Figure 2** shows the location of the solid-liquid interface at different time during the melting process by comparing the experimental data with the numerical result. Euclidean distance between the numerical and experimental results have been calculated. The maximum deviation that occurs at 19 min is less than 4%. The numerical result shows a good agreement with the experimental data. Hence, the numerical method used in the present study can effectively represent the experimentally physical phenomenon [18].

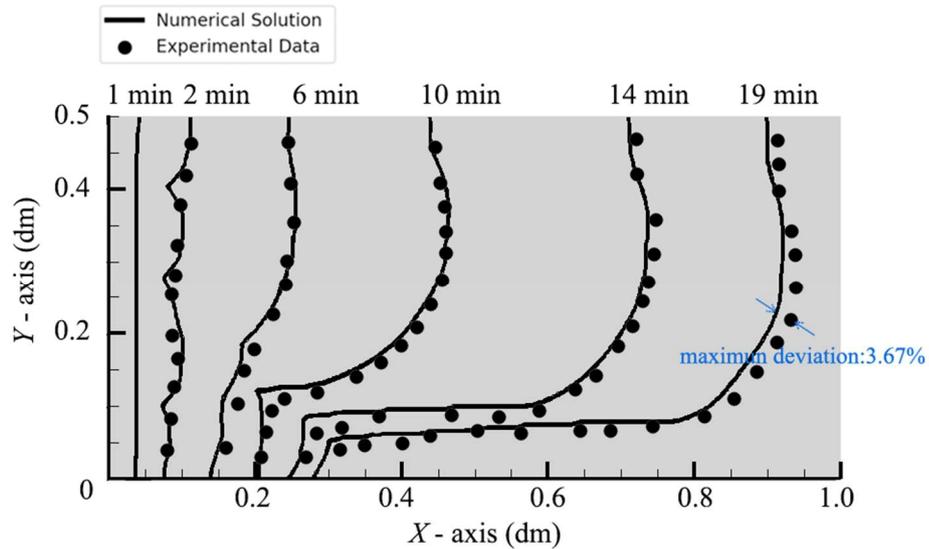

**Fig.2** Numerical and experimental data [42] with the melting front position

In this study, mesh-independent verification was carried out to test the sensitivity of the calculation results to the grids. A variety of mesh schemes are provided in the relevant literature [43], and the number of different grids is verified here. The liquid fraction is the primary reference metric in the present work. When the number of grids is 10,000 and 20,000, the results have deviated, and the maximum deviation is 3% and 1%, respectively. As shown in Fig. 3, when the number of grids reaches 30,000 or more, there is no statistical deviation in the results.

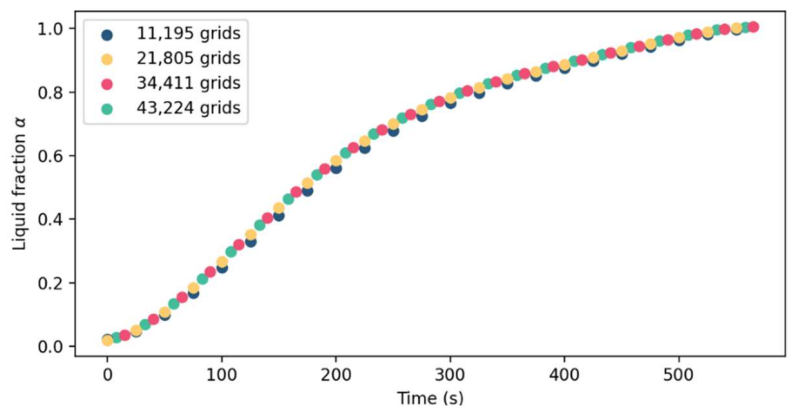

**Fig. 3** Grid-independent verification: liquid fraction of different number of grids

To make the result closer to the real solution, the simulation case calculated in this paper needs to obtain the velocity value at the point position inside the vortex during the charging process. The grid number should be as dense as possible to reduce the discretization error to make the result more accurate. The calculation time of 40,000 grids in the test is within an acceptable range. If the grids exceed 40,000, it not only consumes more calculation time, but leads to problems like rounding error and affects the stability of calculation, resulting in non-convergence in the result. Therefore, under the balance of efficiency and accuracy, the number of grids simulated in this paper is all-around 40,000 grids. Various time steps were used for different cases, and the minimum time step is $1\times10^{-4}$. The computer with a 4.8 GHz AMD Ryzen 9 5900X 12-core processor, which is scalable to 32 Gb of RAM, was used in this study. Each case takes around 1-2 weeks for the complete charging process.

To solve the momentum and energy equations, the SIMPLE method for the pressure-velocity coupling and PRESTO! method for pressure discretization were used. Also, the Least Squares Cell-Based method was adopted for the gradient part. Both momentum and energy discretization used Second-Order Upwind scheme. Besides, the under-relaxation factors for the velocity components, pressure, density, liquid fraction, and energy are 0.7, 1.0, 0.1, 0.9, and 0.9, respectively. The critical dimensionless number of each case in this study is presented in **Table 2,** which contains two cases only consider the natural convection (NC) model. According to **Eqs. (17) and (18)**, $Ra$ is controlled by $L_c$, and $L_c$ and $\Delta T$ control $Ste$. To obtain different melting conditions, $Ra$ and $Ste$ were changed by adjusting the $L_c$ and $\Delta T$. The designed framework of these cases is shown in **Fig.4**.

**Table 2** $Ste$ and $Ra$ of each case

| Case No. | $Ra$ | $Ste$ |
|---|---|---|
| Case 1 | $2.691 \times 10^2$ | 0.103 |
| Case 2 | $2.102 \times 10^3$ | 0.412 |
| Case 3 | $2.102 \times 10^3$ | 0.824 |
| Case 4 | $5.381 \times 10^5$ | 0.206 |
| Case 5 | $1.108 \times 10^6$ | 0.412 |
| Case 6 | $2.691 \times 10^8$ | 0.103 |
| Case 7 (NC) | $3.230 \times 10^6$ | - |
| Case 8 (NC) | $2.691 \times 10^8$ | - |

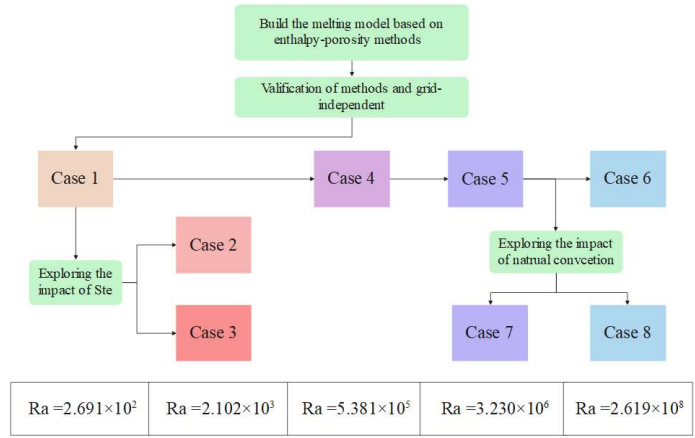

**Fig. 4** Framework of the present study

## 4. Results and discussion

### 4.1 Flow and heat transfer characteristics

An intuitional method to explore nonlinearity is to compare the symmetric degree of the left and right sides of center axis through both quantitively and qualitatively visualization tools. This study used this method as a breakthrough point to investigate the nonlinearity and bifurcation during charging process.

At first, the charging process of paraffin wax is proposed to be divided into three stages in this study, since the melting rate varies significantly during the phase change process. They are the initial stage, the natural convection dominated stage, and the thermal diffusion dominated stage, respectively. The difference in temperature distribution and liquid fraction between stages will be displayed in a visual method. In this section, the critical dimensionless numbers affecting the charging process will be discussed. Hence, different temperature contour maps and flow patterns under various $Ra$ are given to obtain different results.

Cases 1-3 are typically low $Ra$ cases, and **Fig.** gives the charging process in Case 1 for three stages. Due to models' small size, the melting time of this process is short. The pictures in each column represent the exact moment during melting; and the two rows of pictures represent two variables: liquid fraction and temperature. Unlike the temperature distribution, the velocity streamline pattern reveals the state of motion during the charging period, and the liquid fraction reflects the degree of melting. Since there is no flow in the solid zone, the streamline only exists in the liquid and mush zone.

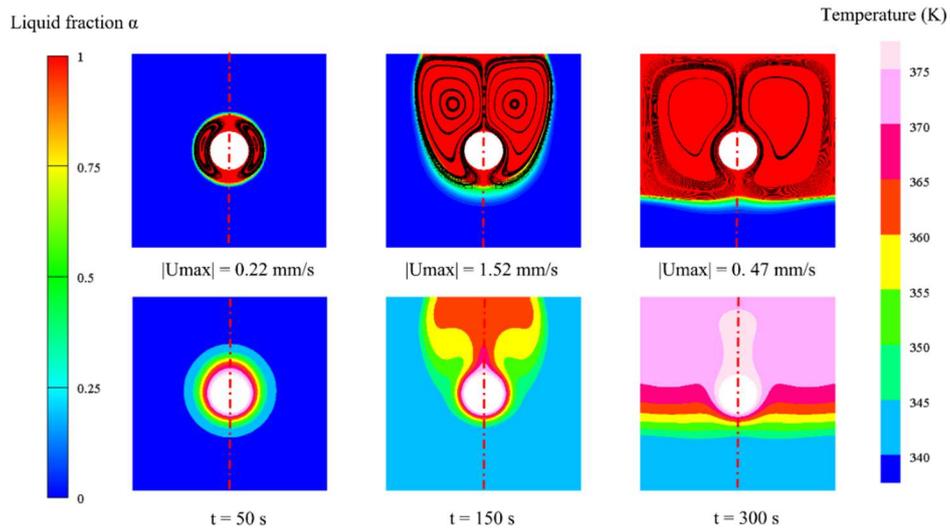

**Fig. 5** Liquid fraction contour maps with flow pattern inside in various time (upside) and temperature distribution in various time (downside) for Case 1 ($2.491 \times 10^2$)

In the initial stage, the flow is weak, and the melting state is average. Then, PCM melts toward the upside to the top of the geometric boundary. At this stage, the melting shape of PCM is symmetrical along with the axis. Eventually, the PCM in the upper space completely melts and the heat diffuses downward regularly.

Cases 2 and 3 were carried out to investigate the impact of *Ste* on the charging process. **Figure 6** shows the contour maps of the liquid fraction. As *Ste* is related to the temperature difference, the charging time is not the same exactly. However, the changing temperature trend is similar, and the whole process is symmetrical along with the axis. Therefore, it is confirmed that *Ste* has a slight effect on the charging process in this study.

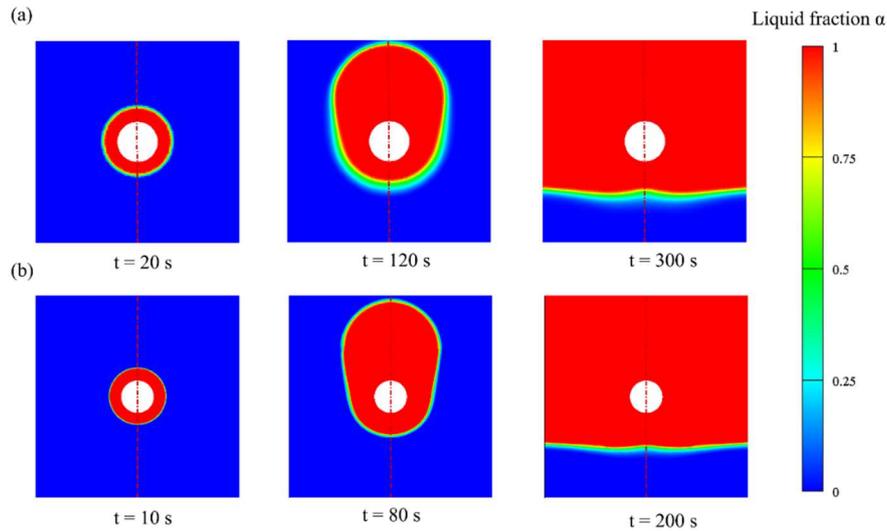

**Fig. 6** Liquid fraction contour maps for Case 2 (a) and Case 3 (b)

**Figure 7** shows the liquid fraction at different melting moments with flow patterns inside during the charging process of Case 4. The initial stage (0 - 10 min) is dominated by the conduction, and the corresponding temperature and liquid fraction contour maps are observed to be symmetric. In the following stage (10 - 300 min), the natural convection dominates, and the melting interface is uneven due to the nonlinearity of natural convection in the charging process. Besides, the main reason for heat convection is that the buoyancy term is considered in the calculation. Then, until the charging process completes (after 300 min), the main form of heat transfer at this stage is conduction. The principal area of the cavity upper the heater has wholly melted after the end of heat exchange stage dominated by convection, and it is difficult for heat convection to transfer to the solid area of PCM, which attributes to the existence of gravity.

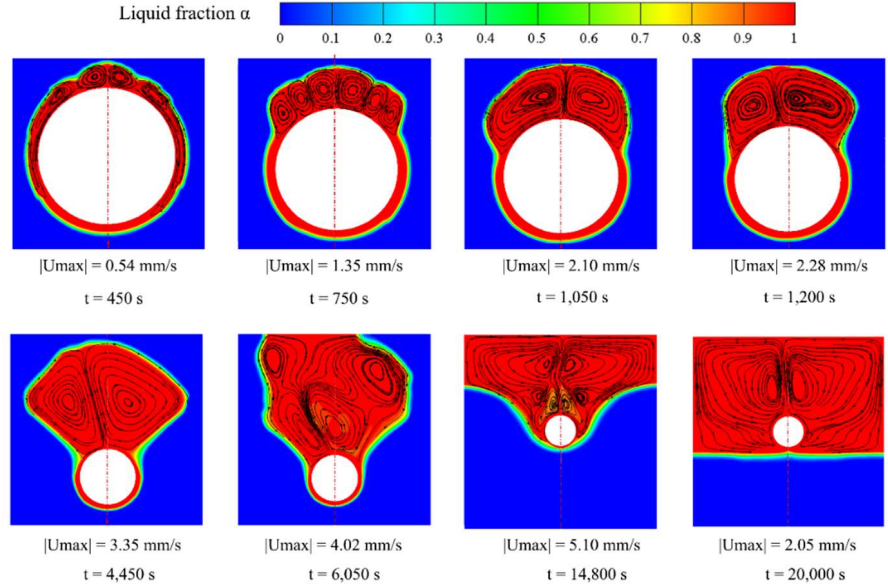

**Fig. 7** Liquid fraction contour maps with flow pattern inside in the exact moments for Case 4

**Figure 8** shows the variation of temperature during the charging process. The observed thermal field shows the temperatures transition of conduction and natural convection in various melting stages. It is worth noting that the temperature distribution is not symmetric along the center axis, and the higher temperature part sways left and right like an indoor flame during the charging process. It is a typical asymmetry in several higher *Ra* cases (Cases 4-7). The asymmetry that appears in the charging process supports the previous experiment studies [44].

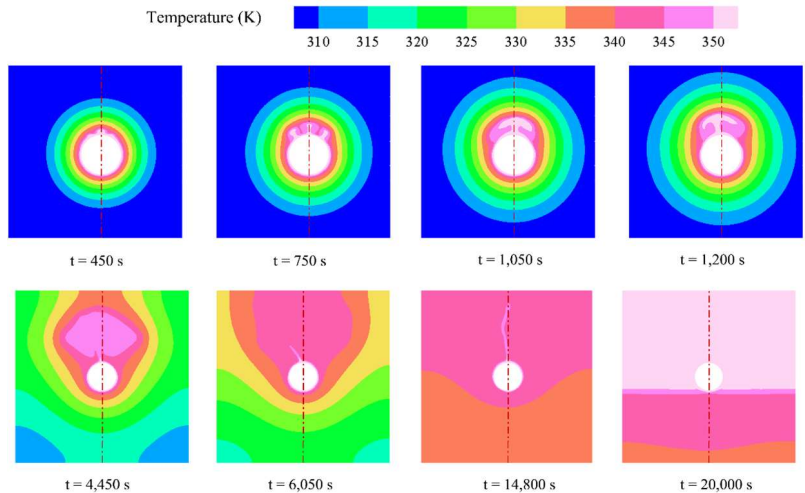

**Fig. 8** Temperature distribution (b) in the exact moments for Case 4

**Figure 7, 8** shows that the flow and heat transfer contour maps in the second stage are evidently asymmetric. This phenomenon is mainly due to the existence of natural convection. In the simulation, the downward gravity has been considered, and the natural convection transfers heat upward. Hence, with the increase of natural convection, two vortices will continuously

squeeze each other along the center axis. This kind of squeeze will have an impact on the melting shape of the PCM eventually, which will be discussed in the following section.

A detailed analysis of the flow and heat transfer behavior during PCM melting is performed to gain deeper insight. **Figure 9** shows the liquid fraction contour maps with velocity streamlines inside for Case 6 at 3,000 s, indicating the evident asymmetric phenomenon. **Figure 9** depicts thermal plumes and multiple counter-rotating vortices formed in the upper space of the heater. The upper unmelted area has a low temperature, whereas the area around the heater has already melted with high temperature. Natural convection will gradually expand from the lower position to the solid boundary of unmelted PCM. The existence of violent heat transfer will eventually form multiple counter-rotating vortices, namely the Benard cell. The classical situation of thermal plumes and unstable Benard cells explain the asymmetry during charging process.

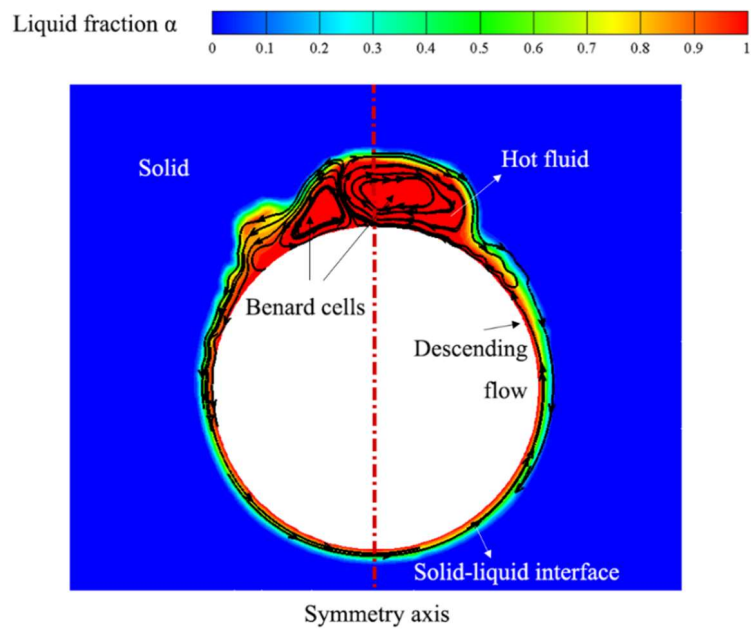

**Fig. 9** Velocity streamlines superimposed on liquid fraction contour map for Case 6 at 3,000 s

To verify the influence of *Ra* and the related buoyancy term at various melting moments, Cases 7 and 8 used pure NC model, i.e., ignored the phase change and heat conduction. The boundary condition at the top is set to a constant temperature to better demonstrate the natural convection. **Figure 10** shows the temperature contour maps under this assumption. The higher part above the heater has extensive similarities with that in **Fig.8**, such changes are more dramatic in **Fig. 10**.

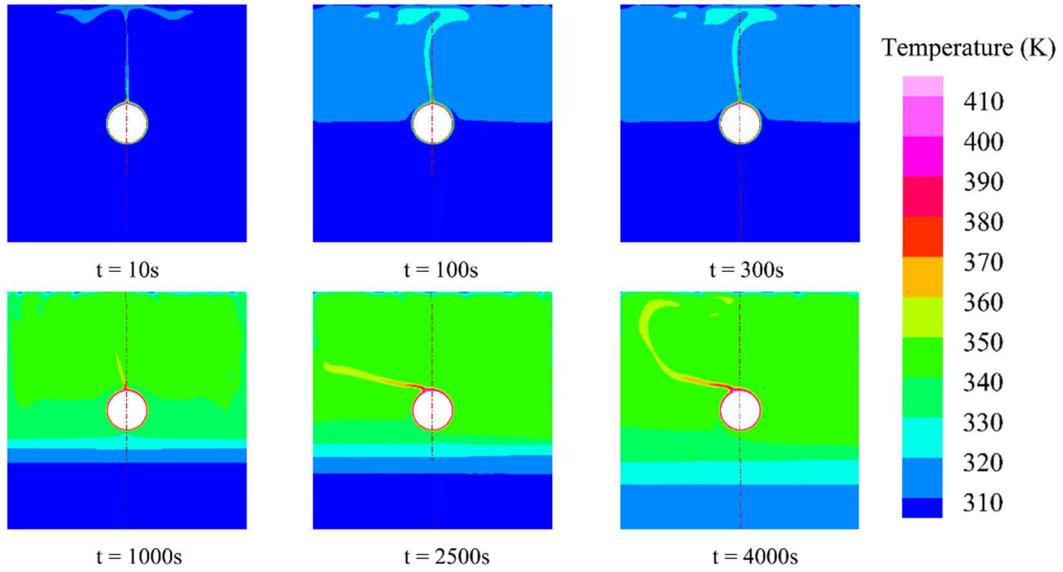

**Fig.10** Temperature distribution at various moments for Case 7 (NC)

The above **Figs. 7-10** visually show the distinct asymmetric phenomenon based on symmetric boundary model. The quantitative degree of symmetry is listed in **Table 3**, which was obtained by extracting the data features of the left and right sides for temperature and liquid fraction during the three-stage charging process. Cases 7 and 8 are the pure NC models without phase change. Therefore, Case 7 in **Table 3** is selected as the representative to assess the symmetry of temperature.

The prominent asymmetries are concentrated in the second stage, i.e., the natural convection dominated stage. The asymmetry means that as *Ra* increases, the numerical results appear multi-solution. Natural convection substantially impacts the nonlinearity of the charging system, which can be found by comparison of Cases 7 and 4. The essence of the asymmetric phenomenon in the liquid fraction contour map is due to the nonlinear characteristics of the charging system itself. A small thermal disturbance during the calculation may affect the stability of the whole system, thus causing multiple solutions in the result. The present study in **Table 3** found that the asymmetric phenomenon begins to appear between Case 3 and Case 4. Thus, the new test cases in section 4.3 were designed based on this finding.

**Table 3** Quantitative asymmetry of charging process in different cases

| Case No. | Degree of symmetry (%) | | | Symmetric state |
|---|---|---|---|---|
| | The first stage | The second stage | The third stage | |
| Case 1 | - | - | - | Symmetry |
| Case 2 | - | - | - | Symmetry |
| Case 3 | - | - | - | Symmetry |
| Case 4 | 98.8 | 93.2 | 96.7 | Asymmetry |
| Case 5 | 98.0 | 89.4 | 99.2 | Asymmetry |
| Case 6 | 94.6 | 91.3 | 98.0 | Asymmetry |
| Case 7 (NC) | | 92.1 | | Asymmetry |

## 4.2 Nonlinear characteristics

In this section, the nonlinear characteristics during the charging process were discussed. Various dynamic analysis methods such as time series and phase portraits were used. Due to the high viscosity of paraffin wax, even the maximum velocity of paraffin wax in the charging process is negligible. Therefore, the dimensionless velocities were used in both the time series and phase portraits. These data are all raw data obtained through numerical calculations without filtering or fitting processing, objectively reflecting the system's nonlinearity. The phase portrait diagram reflects the stability of the system's flow and heat transfer at the monitoring point during the melting process. When the phase portrait diagram presents a simple closed figure, it reflects that the flow state tends to be stable. And when the phase portrait enters an unclosed or complex disordered pattern, it shows that the monitoring point locates in a transition position and becomes unsteady.

**Figure 11** shows time-series and phase portraits of velocity for two cases (Cases 1 and 2) at the same monitor point. To better show the nonlinearity and oscillation, time series were partially magnified. It is found from **Figs.11(a) and (c)** that the charging time is longer, and the melting rate is higher for Case 2. **Figure 11** does not include the steady-state area where the velocity is stable until completely melted (the final velocity is zero). The velocity oscillation mainly occurs in the initial melting stage and gradually weakens until the end of melting. Phase portraits are relatively regular and straightforward, but the graph is not closed for Case 2 as shown in **Fig.11(d)**.

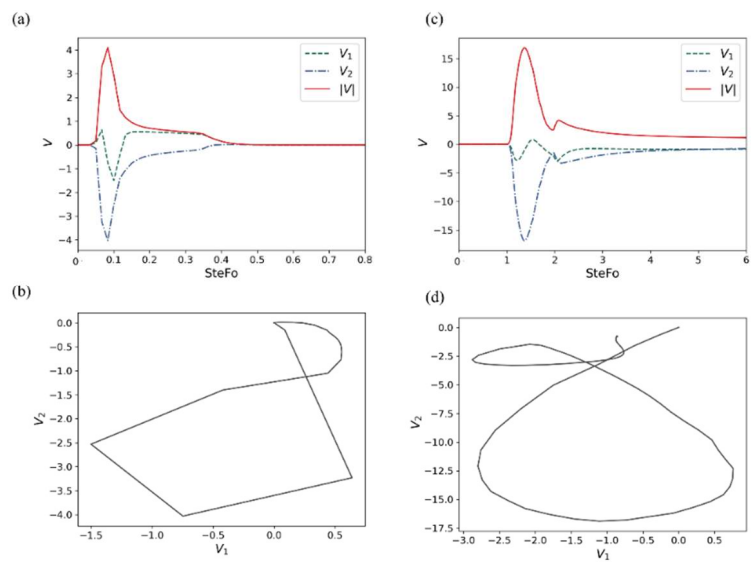

**Fig.11** Time-series and phase portraits of velocity for Case 1 (a and b) and Case 2 (c and d)

**Figure 12** shows time-series and phase portraits of velocity for Case 3. These data were obtained from the monitoring point. The velocity of the $X$-component changes drastically and the curve is more oscillating compared with that of the $Y$-component. The overall phase portraits fluctuate, but the orbits are smooth. Compared with **Fig.11**, the difference of time series is in an acceptable range. The phase portraits of velocity at low *Ra* maintain relatively smooth orbits.

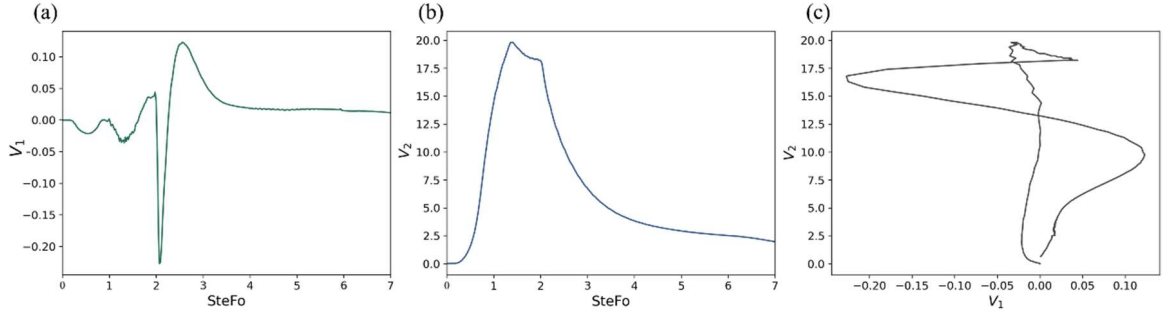

**Fig.12** Time-series (a and b) and phase portraits (c) of velocity for Case 3

**Figure 13** shows time series and phase portraits of velocity for Case 8 (NC) at the same position of Case 1. It is observed that the time-series of velocity in Case 8 is in drastic oscillation from time 0 s. The velocity change of *X*-component is relatively concentrated, and the apparent rising stage of *Y*-component can be found in **Fig.13(b)**. The decrease of the absolute velocity indicates that the direction of the velocity is the negative *Y*-component and the oscillation interval is more remarkable than that of *X*-component (V1). Phase portrait shows a chaotic state, and the orbits of velocity are not sharpened. Attractor reconstruction from the time series of velocities can be found in **Fig.13(c)**.

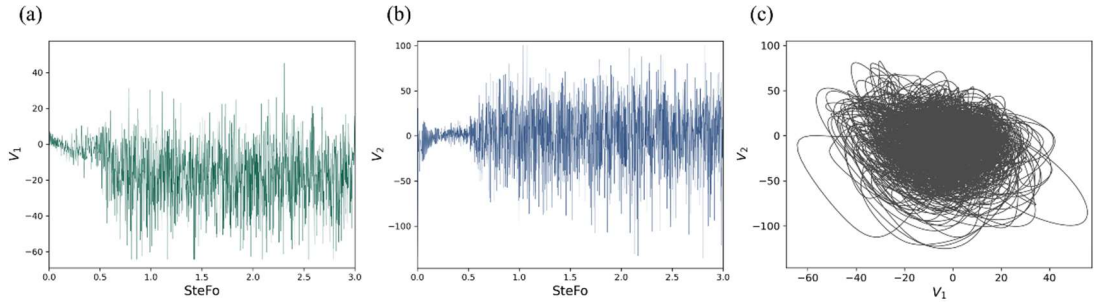

**Fig.13** Time-series (a and b) and phase portraits (c) of velocity for Case 8 (NC)

**Figure 14** shows the chaotic behavior for Case 4 during the charging process. Since the monitor position is in the solid region of PCM from time 0 s, the velocity variation is zero at the beginning. When the melting begins, the oscillation interval of velocity gradually increases with time. Based on the analysis of charging process, oscillation change can be explained as follows. At the initial stage, the velocity suddenly appears, at this time the heat transfer is mainly dominated by conduction, the flow in the liquid phase is not apparent, and the oscillation interval of velocity is small. Then, natural convection occupies the dominant status along with the appearance of Benard cells in the flow, and the oscillation starts to intensify. Furthermore, from these time-series, the velocity in both directions remains at the unsteady state. When the *Ra* is higher than $5.381 \times 10^5$, the time-series and phase portraits of velocity in all cases are similar to which in the NC case in **Fig. 13**. The phase portraits are concentrated into a cluster, which shows irregular changes by comparing **Fig. 12(c)** with **Fig. 14(c)**. Therefore, it is concluded that natural convection leads to nonlinearity, and the system is in an unsteady state.

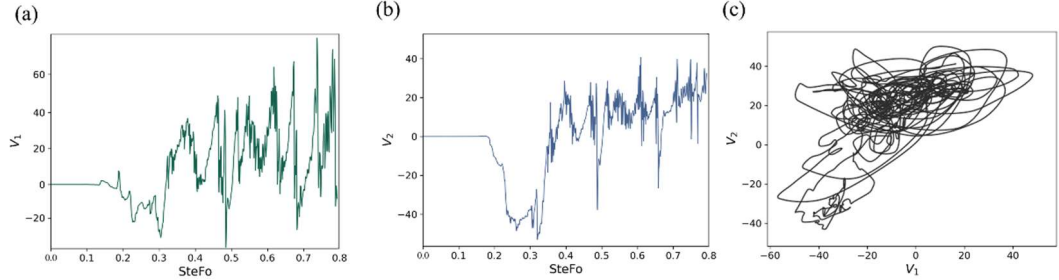

**Fig. 14** Time-series (a and b) and phase portraits (c) of velocity for Case 4

**Figure 15** shows time-series and phase portraits of velocity in the initial charging process for Case 6. The time series was enlarged in the graph to highlight results. Compared to the velocity of *Y*-component (V2), the *X*-component (V1) oscillates more violently and increases gradually for the velocity. The appearance of this change is different from that in the time series in **Fig. 14** (Case 4), especially in the speed difference of the *Y*-component. Unlike the phase portraits in the former cases, the orbits' adjacent distance gradually increases with ellipse shape.

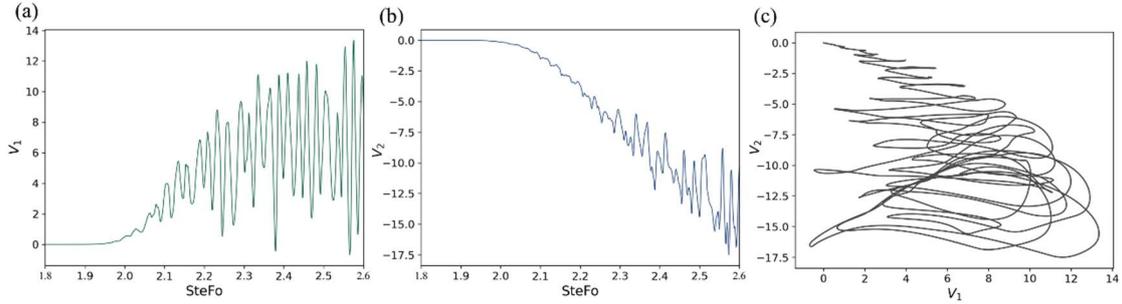

**Fig. 15** Time-series (a and b) and phase portraits (c) of velocity for Case 6

By drawing the time series and phase portraits utilizing nonlinear dynamics to analyze the charging process, nonlinear rules of the system can be obtained in the present study. More importantly, the exploration lies on the phase orbits, which enables the rule of the flow state to be more accurate and intuitive to show the nonlinear characteristics. At a low *Ra* melting, the flow is relatively stable, only one or two orbits are found in the phase portrait. However, the different orbits overlap as the *Ra* value increases, and the flow is gradually chaotic. Under the model that only considers natural convection, the correspondence between each orbit in the phase portrait is clear. These orbits develop from small to large, reflecting that the natural convection has a more noticeable impact on the system's nonlinearity.

Based on the discussion of asymmetric phenomena and results of cases, a specific nonlinear law is found about the nonlinearity of charging process varying with $Ra$ value. The flow stability is displayed through the time series of velocity and phase portraits. Flow and heat transfer are stable when $Ra < 2.102 \times 10^3$ (**Fig. 5** and **Fig.11**), begin to transit to unsteady state when $Ra > 1.345 \times 10^5$ (**Fig. 7, 8** and **Fig.14** ), and begin to oscillate vigorously and irregularly when $Ra > 2.691 \times 10^8$ (**Fig. 9** and **Fig.15**). The variation of $Ra$ indicates a bifurcation point in mathematics, which is the key of charging process changing from one single solution to multiple solutions.

**4.3 Comparison of half and entire domain models**

The above discussed mechanism of the asymmetric phenomenon and flow stability for charging process is based on the adopted entire domain model, of which the advantages over the half domain model are unknown. Therefore, the two models are compared in this section with their description of energy storage features concerned in applications. This section's simulation is based on the results in **Table 3** that when Ra > 1×10$^4$, the simulation results show obvious asymmetric phenomenon. Therefore, the present numerical experiments were set within this Ra range using the half and entire domain models. The difference of energy storage capacity represented by liquid fraction is shown in **Fig. 16**.

It is found that the calculated liquid fraction of half domain model is larger than that of entire domain model throughout the charging process. Especially when the thermal energy storage capacity equals to 350 kJ, the extreme condition occurs with a maximum relative deviation of 9.24%, which is intolerable to the precise control of energy storage system. At this moment, the interphase boundary is asymmetric, seen from the result of entire domain model calculation; however, this asymmetry cannot be revealed by the half domain model calculation. Therefore, further study and application need to consider the asymmetric phenomenon and to use the entire domain model.

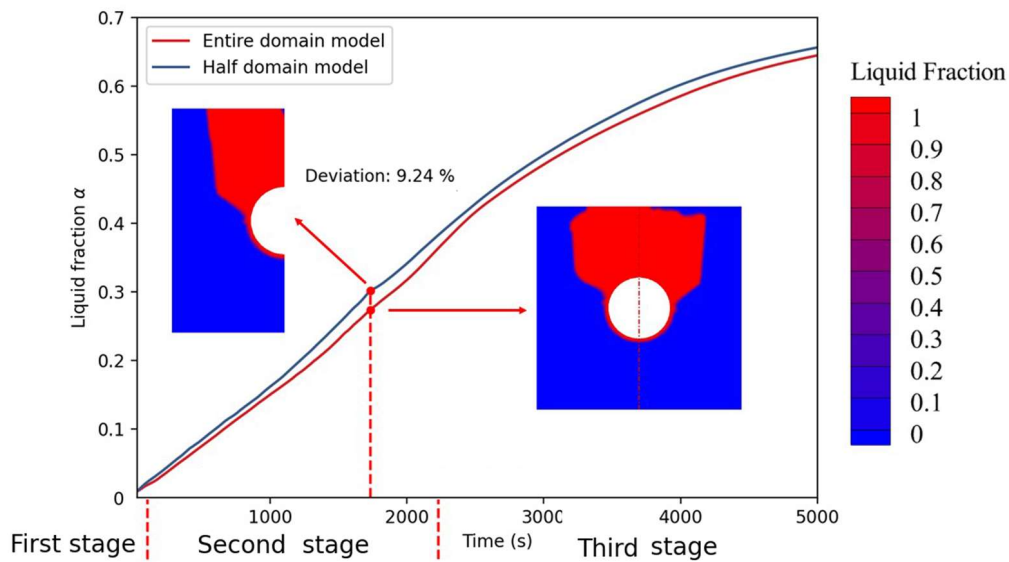

**Fig. 16** Difference of energy storage capacity calculation between entire and half domain models

The comparison of charging speed calculated by the entire and half domain models is shown in **Fig. 17**. The average value of liquid fraction changes every five seconds was used to obtain the change rate of solid and liquid volume and was converted to the charging speed.

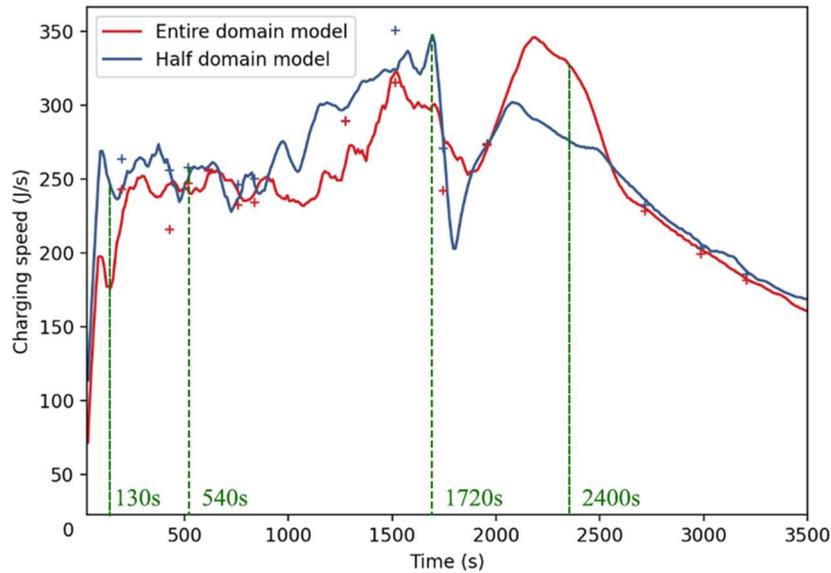

**Fig. 17** Difference of charging speed calculation of entire and half domain models

In the early melting stage, the charging speed of the half domain model increases faster. Subsequently, the charging speed calculated by the two models keeps fluctuating around 250 J/s from 250 s to 1,000 s. The number of vortices inside is typically equal at this time, as shown in **Fig.18**. With the development of charging process, the melting in the half domain model is gradually intense, and the internal vortexes swallow and squeeze each other. And then, the overall charging speed of the half domain model is higher than that of the entire domain one.

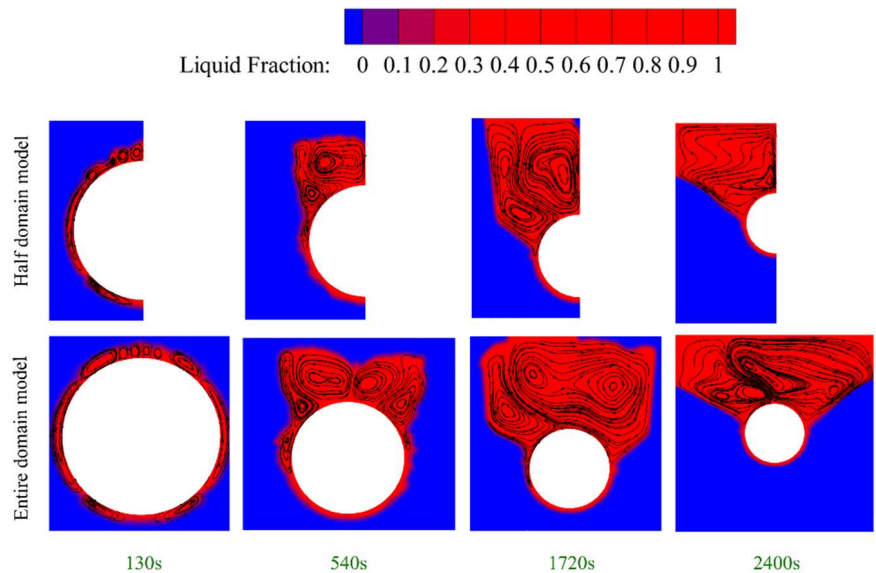

**Fig. 18** Comparison of the liquid fraction contour with flow pattern inside for entire and half

domain models at four typical moments

Around 1,600 s, both models show a significant decrease in the charging speed because the existence of the upper boundary wall blocks the charging process. However, the left liquid zone of the entire domain model is not completely symmetrical to the right side. When a part of the liquid zone melts to the upper wall, the rest of the upper part is still in a state of solid, which impacts the charging speed of the entire domain model. The difference of contours for temperature and liquid fraction with flow pattern calculated by the entire and half domain models are shown in **Fig. 19**. The melting flow and heat transfer state were captured at 1,720 s.

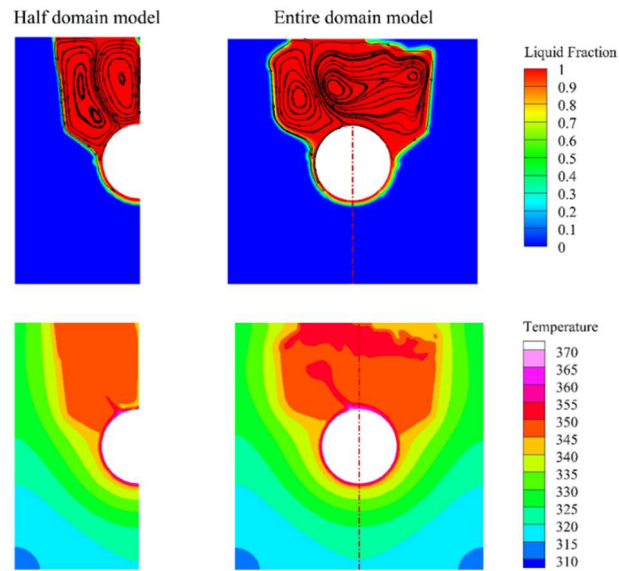

**Fig. 19** The liquid fraction contour with flow pattern inside and temperature contour at 1,720 s

When enters the third stage of charging process, the melting begins to diffuse from the middle to both sides, and the charging speeds show a fast upward trend. The liquid zone on the left and right sides of the entire domain model is asymmetric, as shown in **Fig. 18**. It is seen that the flow of vortices in the entire domain model is more vigorous and evolves more vortices from the last time. By contrast, the internal flow in the half domain model is more moderate. Therefore, the melting rate has a significant increase, as shown in **Fig. 17**.

It is obvious that the flow and heat transfer are asymmetric, seen from both the liquid fraction and temperature contours using the entire domain model. However, the asymmetric behavior cannot be recognized using only a half domain model. What's worse, if the half domain model was used, there would be three more vortexes at the uncalculated right side based on inertial thinking. In fact, only three vortexes are found in total under the same condition using the entire domain model. The number of vortexes reflects the flow strength; thus, the flow in the half domain model may be misunderstood as over fierce, e.g., the over higher charging speed. This study concludes that the asymmetric phenomenon is influenced by multi-solution, which is typical in numerical calculation of melting process.

**4.4 Energy storage performance analysis**

The significance of asymmetric phenomenon to energy storage application will be discussed in this section by analyzing the performance of charging speed and thermal energy storage capacity. A quantitative method was proposed and used to express the values of nonlinear characteristics. The liquid fraction data on both sides of the symmetry axis at a certain moment were extracted as a data feature to compare the symmetric degree, given by the percentage of the absolute difference of liquid fractions at both sides over the liquid fraction at left side. This process was automated by building a Python program.

**Figure 20** shows the symmetric degree during the charging process along with liquid fraction. In addition, the symmetric degree at the different moment is also shown in the figure. The symmetric degree decreases quickly once the charging begins, though the asymmetry is not apparent at the initial stage (< 2%). At the second stage, the asymmetry increases to the most remarkable situation (> 14%); and then the symmetry is gradually recovered. The asymmetry reduces to about 1% at the end of the third stage. The asymmetry mainly occurs at the first half of charging process. Typical asymmetry is found with the liquid fraction of < 0.5, i.e., the natural convection dominated stage.

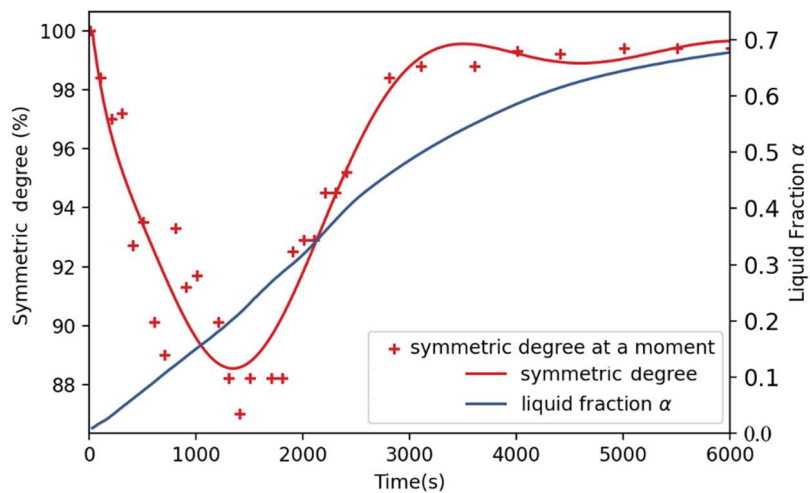

**Fig.20** Symmetric degree change with liquid fraction

The charging speed in **Fig. 17** has an inverse development trend compared with the symmetric degree in **Fig. 20**. When the charging speed is higher than 245 J/s, the flow in liquid zone changes drastically, resulting in increasing nonlinearity. Around 1,800 s, an abnormal drop of charging speed is found, due to the top wall of the model boundary blocks the upper part of PCM and brings a substantial fluctuation in the overall charging speed.

Potential measures to enhance the charging speed proposed by this study is to increase the nonlinearity of flow and heat transfer, according to the relationship between asymmetry and charging speed. One measure is adding thermal disturbance during the melting process to disrupt the stable state of heat flow and to enhance the convective heat transfer.

The thermal energy stored in PCM includes the latent and sensible heat, computed by Eq. (19).

$$Q_{PCM} = m(c_s(T_m - T_{ini}) + \Delta h + c_l(T_h - T_m)), \qquad (19)$$

where $c_s$ and $c_l$ are the specific heat of PCM at solid and liquid states, respectively.

Four scenarios were simulated with the energy storage capacity of 232.5, 256.0, 350.0, and 444.0 kJ, of which the average symmetric degrees of charging process are shown in **Fig. 21**. It is observed that the thermal energy storage capacity also has a positive correlation with asymmetry, i.e., a negative correlation with symmetry degree. And calculating the average value of symmetric degree for the four scenarios, a clearer change rule is shown. The overall symmetric degree gradually decreases as the energy storage capacity increases. According to Eq. 19, the thermal capacity is positively related to the temperature difference and specific heat capacity of liquid. Meanwhile, the Rayleigh number definition (Eq. 17) reflects that the temperature difference and specific heat capacity of liquid is proportional to Ra. As the specific heat capacity of liquid is constant, it is the temperature difference that dominants the synchronous change of thermal energy storage capacity and Ra, as well as the asymmetry. It is concluded that the asymmetry behavior varies with the capacity of energy storage system, and the higher the capacity, the greater the asymmetry. A possible worse situation is that the calculation deviation of half domain model may become larger than 10% (the example in **Fig. 16**) for higher energy storage capacity scenarios.

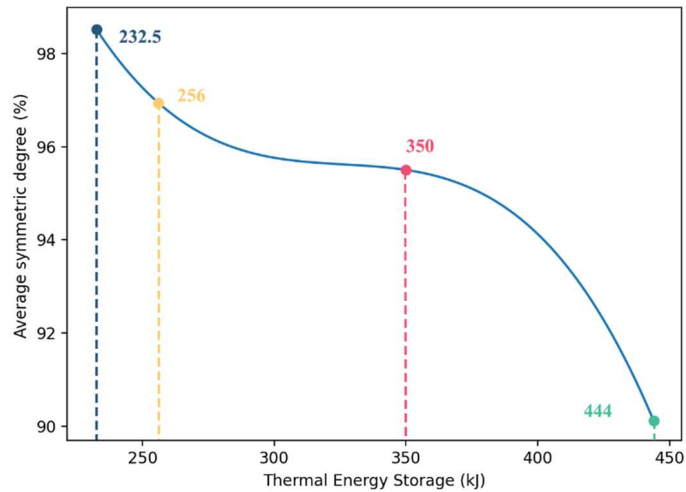

**Fig. 21** Symmetric degree of charging processes with different thermal energy storage capacity

In addition, the system with higher capacity is more prone to have multiple solutions, i.e., the asymmetric phenomenon. Therefore, in the thermal energy storage system with higher capacity, the bifurcation of multi-solution cannot be avoided, and the entire domain model is more necessary to adopt. Calculating as many times as possible is also needed to get sufficient and accurate solutions in LHTES application. This work has proved the evident importance of asymmetric phenomenon to energy storage. Especially, it is found that the asymmetry is positively correlated with the thermal capacity. For phase change in large size domains with higher thermal capacity, the asymmetric phenomenon may be more considerable and non-negligible to the energy storage system. The qualitative conclusions on the nonlinearity of LHTES are general to the same type of problems. Whereas for the quantitative features, e.g., the asymmetric degree calculated in the present study, further particular treatment is needed for models with different structures.

## 5. Conclusions

This study sets out to determine the asymmetric phenomenon in the charging process of phase change energy storage by numerical simulation. The results calculated by entire domain model give a reliable description of the charging process of paraffin wax as PCM. The visualization method was conducted to illustrate the nonlinearity in various cases. Energy storage performance is discussed with the symmetric degree influenced by nonlinearity. It is of reference significance for numerical investigation in the related works about phase change process. The following conclusions can be obtained.

1. The asymmetric phenomenon during the charging process of phase change energy storage has been revealed. The visualization method finds that the temperature and liquid fraction contours and the streamline pattern are all asymmetric. The charging process is divided into three stages by different domination types of heat transfer.

2. Rayleigh number analysis indicates a bifurcation point, and three important Rayleigh numbers are found. Flow and heat transfer are stable when $Ra < 2.102 \times 10^3$, begin to transit to unsteady state when $Ra > 1.345 \times 10^5$, and begin to oscillate vigorously with an irregular and chaotic phase portrait when $Ra > 2.691 \times 10^8$.

3. The numerical results with half domain and entire domain models are evidently different. Deviations may be larger than 10% for typical asymmetric flow and heat transfer. Further study and application need to consider the bifurcation and to use the entire domain model to accurately calculate the charging process. The appearance of multi-solution is confirmed, there are more than one solution for charging process.

4. The asymmetry behavior varies with the capacity of thermal energy storage system; and the higher the capacity, the greater the asymmetry. The charging speed has a positive correlation with asymmetry. Potential measures to enhance the charging speed is to increase the nonlinearity of flow and heat transfer by destroying the flow stability and enhancing the convective heat transfer.

5. The results obtained from the calculation are multiple solutions, which will differently impact the engineering application of LHTES. It is necessary to consider the nonlinearity; specifically, nonlinear LHTES engineering application includes two aspects. The one is calculating as many results as possible to help analyze the physical process and achieve precise system control. The other is making the engineering state enter the state under the optimal numerical solution by enhancing asymmetry, e.g., adding thermal disturbance.

## Acknowledgments

The work described in this paper was supported by the National Natural Science Foundation of China through grant No. 51736007. This work was also Sponsored by Shanghai Sailing Program (No. 19YF1434800).

## Nomenclatures

| **Abbreviation** | | $c$ | specific heat (kJ/(kg·K)) |
|---|---|---|---|
| PCM | phase change material | $S_m$ | Darcy type source term (kg/(m²·s²)) |
| LHTES | latent heat thermal energy storage | $S_b$ | buoyancy source term (kg/(m²·s²)) |
| NC | natural convection | $S_M$ | dimensionless Darcy type source term |
| | | $S_B$ | dimensionless buoyancy source term |
| **Symbols** | | | |
| $Ra$ | Rayleigh number | **Greek letter** | |
| $Ste$ | Stefan number | $\alpha$ | liquid fraction |
| $Fo$ | Fourier number | $\beta$ | thermal expansion coefficient of liquid (1/K) |
| $Pr$ | Prandtl number | $\lambda$ | thermal conduction (W/(m·K)) |
| $L_c$ | characteristic length (m) | $\theta$ | dimensionless of excess temperature |
| $v$ | velocity vectors (m/s) | $\rho$ | density (kg/m³) |
| $g$ | gravitational acceleration (m/s²) | $\nu$ | kinematic viscosity (m²/s) |
| $T$ | temperature (K) | | |
| $p$ | pressure (Pa) | **Subscripts** | |
| $P$ | dimensionless pressure | ini | initial |
| $v$ | velocity vectors (m/s) | sen | sensible |
| $V$ | dimensionless velocity vectors | s | solidus |
| $t$ | time (s) | l | liquidus |
| $h$ | specific enthalpy (kJ/kg) | m | melting |
| $H$ | dimensionless enthalpy | ref | reference |
| $\Delta h$ | latent heat (kJ/kg) | h | heater |